\documentclass[twocolumn,twoside,preprintnumbers,amsmath,amssymb,pacs]{revtex4}
\usepackage{epsfig}
\usepackage{graphicx}% Include figure files
\usepackage{dcolumn}% Align table columns on decimal point
\usepackage{bm}% bold math
\newcommand{\lms}{\Lambda_{\overline{\mbox{\tiny{MS}}}}}

\usepackage{pslatex}

\begin{document}
\preprint{LTH-695}
\title{{\Large  The infrared behavior of the gluon and ghost propagators\\ in SU(2) Yang-Mills
theory in the maximal Abelian gauge\footnote{Talk given by S.P.
Sorella at the ``I Latin American Workshop on High Energy
Phenomenology(I LAWHEP)'', December 1-3 2005, Instituto de
F\'{i}sica, UFRGS, Porto Alegre, Rio Grande Do Sul, Brasil}
 }}

%\preprint{LTH-695 }
\author{M.A.L. Capri$^a$}
\email{marcio@dft.if.uerj.br}
\author{D. Dudal$^b$}\email{david.dudal@ugent.be}
\author{J.A. Gracey$^{c}$}
    \email{jag@amtp.liv.ac.uk}
\author{V.E.R. Lemes$^{c}$}\email{vitor@dft.if.uerj.br}
\author{R.F. Sobreiro$^a$}
 \email{sobreiro@dft.if.uerj.br}
\author{S.P. Sorella$^a$}
\email{sorella@uerj.br} \altaffiliation{Work supported by FAPERJ,
Funda{\c c}{\~a}o de Amparo {\`a} Pesquisa do Estado do Rio de
Janeiro, under the program {\it Cientista do Nosso Estado},
E-26/151.947/2004.}
\author{R. Thibes$^a$}\email{thibes@dft.if.uerj.br}
\author{H. Verschelde$^a$}
 \email{henri.verschelde@ugent.be}
 \affiliation{\vskip 0.1cm $^a$ UERJ - Universidade do Estado do Rio de
Janeiro\\Rua S\~{a}o Francisco Xavier 524, 20550-013
Maracan\~{a}\\Rio de Janeiro, Brasil\\\\\vskip 0.1cm $^b$ Ghent
University
\\ Department of Mathematical
Physics and Astronomy \\ Krijgslaan 281-S9 \\ B-9000 Gent,
Belgium\\\\
\vskip 0.1cm $^c$ Theoretical Physics Division\\ Department of
Mathematical Sciences\\ University of Liverpool\\ P.O. Box 147,
Liverpool, L69 3BX, United Kingdom }

%\received{on 25 March, 2006}

\begin{abstract}
We report on some recent analytical results on the behaviour of the
gluon and ghost propagators in Euclidean $SU(2)$ Yang-Mills theory
quantized in the maximal Abelian gauge (MAG). This gauge is of
particular interest for the dual superconductivity picture to
explain color confinement. Two kinds of effects are taken into
account: those arising from a treatment of Gribov copies in the MAG
and those arising from a dynamical mass originating in a dimension
two gluon condensate. The diagonal component of the gluon propagator
displays the typical Gribov-type behaviour, while the off-diagonal
component is of the Yukawa type due to the dynamical mass. These
results are in qualitative agreement with available lattice data on
the gluon propagators. The off-diagonal ghost propagator exhibits an
infrared enhancement due to the Gribov restriction, while the
diagonal one remains unaffected.\newline\newline PACS numbers:
11.15.-q,11.15.-Tk\newline\newline Keyword: Yang-Mills gauge theory;
maximal Abelian gauge; Faddeev-Popov operator; propagators;
Gribov-Zwanziger scenario; condensates
\end{abstract}

\maketitle

\setcounter{page}{0}

\section{Introduction}
We shall consider pure Euclidean $SU(N)$ Yang-Mills theories with
action
\begin{equation}\label{1}
{ S_{YM}}=\frac{1}{4}\int d^4 x\;\mathcal{F}_{\mu \nu
}^{a}\mathcal{F}^{a}_{\mu \nu }\;,
\end{equation}
where $\mathcal{A}_{\mu }^{a}$, $a=1,...,N^{2}-1$ is the gauge boson
field, with associated field strength
\begin{equation}\label{3}
\mathcal{F}^{a}_{\mu \nu }=\partial _{\mu }\mathcal{A}_{\nu
}^{a}-\partial _{\nu }\mathcal{A}_{\mu
}^{a}+gf^{abc}\mathcal{A}_{\mu }^{b}\mathcal{A}_{\nu }^{c}\;.
\end{equation}
The theory (\ref{1}) is invariant w.r.t. the local gauge
transformations
\begin{equation}\label{4}
    \delta \mathcal{A}_\mu^a = \mathcal{D}_{\mu}^{ab}\omega^b\;,
\end{equation}
with
\begin{equation}\label{5}
    \mathcal{D}_{\mu}^{ab}=\partial_\mu\delta^{ab}-gf^{abc}\mathcal{A}_\mu^c\;,
\end{equation}
denoting the adjoint covariant derivative.

As it is well known, the theory (\ref{1}) is asymptotically free
\cite{Gross:1973id, Politzer:1973fx}, i.e. the coupling becomes
smaller at lower energies and vice versa. At very high energies, the
interaction is weak and the gluons can be considered as almost free
particles. However, in spite of the progress in the last decades, we
still lack a satisfactory understanding of the behaviour of
Yang-Mills theories in the low energy regime. Here the coupling
constant of the theory is large and nonperturbative effects have to
be taken into account. One of the greatest challenges of
contemporary theoretical physics is to actually prove that (\ref{1})
describes a \emph{confining} theory in the infrared. Its physical
spectrum is believed to be given by colorless bound states of gluons
called glueballs. We adopt here the common paradigm according to
which the first step for understanding color confinement in the real
world, i.e. in the QCD-world,  is to understand it first in pure
Yang-Mills theory (\ref{1}).

A partial list of causes of nonperturbative effects is given by
Gribov ambiguities which affect the Faddeev-Popov quantization
procedure and hence the propagators \cite{Gribov:1977wm}, the
existence of condensates such as $ \langle F^2 \rangle$, $\langle
A^2 \rangle$, etc.
\cite{Shifman:1978bw,Gubarev:2000eu,Dudal:2003by}, the existence of
(topologically) nontrivial field configurations like instantons
\cite{Schafer:1996wv}, etc. These effects are not necessarily
unrelated, as e.g. instantons can contribute to the condensate $
\langle F^2 \rangle$ \cite{Schafer:1996wv}, the Faddeev-Popov
operator has zero modes in an instanton background
\cite{Maas:2005qt}, etc.

\subsection{The dual superconductivity picture}
A particularly appealing proposal to explain confinement in the low
energy regime was made in
\cite{thooft,Nambu:1974zg,Mandelstam:1974pi}. Let us give a simple
idea of the picture. If the QCD vacuum would contain magnetic
monopoles and if these monopoles would condense, there will be a
dual Meissner effect which squeezes the chromoelectric field into a
thin flux tube. This results in a linearly rising potential,
$V(r)=\sigma r$, between static charges, as can be guessed from
Gauss' law, $\int EdS=const$ or, since the main contribution is
coming from the flux tube, one finds $E\Delta S\approx const$, hence
$V=-\int Edr\propto r$ where $\Delta S$ is the area of the flux
tube. In fact, it is not difficult to imagine the longer the flux
tube (string) gets, the more energy it will carry. Hence, it would
cost enormous amounts of energies to separate two test charges from
each other and so they are confined to stay together.

\subsection{The maximal Abelian gauge} An important ingredient for
the dual superconductivity picture to work, is clearly the  presence
of (magnetic) monopoles. As the Yang-Mills action (\ref{1}) is
locally gauge invariant, we need to fix the gauge in order to
quantize it. 't Hooft invented the so-called Abelian gauges in
\cite{'tHooft:1981ht}, of which the maximal Abelian gauge (MAG) is a
specific example
\cite{'tHooft:1981ht,Kronfeld:1987ri,Kronfeld:1987vd}. In these
Abelian gauges, defects in the gauge fixing give rise to point-like
singularities which can be associated to monopoles.

We may decompose the $SU(2)$ gauge field into off-diagonal and
diagonal components, according to
\begin{equation}
\mathcal{A}_{\mu }=A_{\mu }^{a}T^{a}+A_{\mu }T^{3}\;,  \label{conn}
\end{equation}
where $T^{a}$, $a=1,2$, denote the off-diagonal generators of
$SU(2)$, while $T^{3}$ stands for the diagonal generator,
\begin{eqnarray}
\left[ T^{a},T^{b}\right] &=&i\varepsilon
^{ab}T^{3}\;,\;\;\;\;\left[ T^{3},T^{a}\right] =i\varepsilon
^{ab}T^{b}, \label{la1}
\end{eqnarray}
where
\begin{eqnarray}
\varepsilon ^{ab} &=&\varepsilon ^{ab3}\;,\;\;\;\; \varepsilon
^{ac}\varepsilon ^{ad} =\delta ^{cd}\;.  \label{la2}
\end{eqnarray}
Similarly, for the field strength one shall find
\begin{equation}
\mathcal{F}_{\mu \nu }=F_{\mu \nu }^{a}T^{a}+F_{\mu \nu }T^{3}\;,
\label{fs}
\end{equation}
with off-diagonal and diagonal parts given by
\begin{eqnarray}
F_{\mu \nu }^{a} &=&D_{\mu }^{ab}A_{\nu }^{b}-D_{\nu }^{ab}A_{\mu
}^{b}\;,
\label{fsc} \\
F_{\mu \nu } &=&\partial _{\mu }A_{\nu }-\partial _{\nu }A_{\mu
}+g\varepsilon ^{ab}A_{\mu }^{a}A_{\nu }^{b}\;,  \nonumber
\end{eqnarray}
The covariant derivative $D_{\mu }^{ab}$ is defined with respect to
the diagonal component $A_{\mu }$
\begin{equation}
D_{\mu }^{ab}\equiv \partial _{\mu }\delta ^{ab}-g\varepsilon
^{ab}A_{\mu }\;.  \label{cv}
\end{equation}
For the Yang-Mills action in Euclidean space, one finally obtains
\begin{equation}
S_{\mathrm{YM}}=\frac{1}{4}\int d^{4}x\,\left( F_{\mu \nu
}^{a}F_{\mu \nu }^{a}+F_{\mu \nu }F_{\mu \nu }\right) \;.
\label{ym}
\end{equation}
As it is easily checked, the classical action (\ref{ym}) is left
invariant by the decomposed gauge transformations
\begin{eqnarray}
\delta A_{\mu }^{a} &=&-D_{\mu }^{ab}{\omega }^{b}-g\varepsilon
^{ab}A_{\mu
}^{b}\omega \;,  \nonumber \\
\delta A_{\mu } &=&-\partial _{\mu }{\omega }-g\varepsilon
^{ab}A_{\mu }^{a}\omega ^{b}\;.  \label{gauge}
\end{eqnarray}
The MAG is obtained by demanding
\begin{equation}
D_{\mu }^{ab}A_{\mu }^{b}=0\;,  \label{offgauge}
\end{equation}
which follows by requiring that the auxiliary functional
\begin{equation}
\mathcal{R}[A]=\int {d^{4}x}A_{\mu }^{a}A_{\mu }^{a}\;,
\label{fmag}
\end{equation}
is stationary with respect to the gauge transformations
(\ref{gauge}).

Moreover, as it is apparent from the presence of the covariant derivative $%
D_{\mu }^{ab}$, (\ref{offgauge}) allows for a residual local $U(1)$
invariance corresponding to the diagonal subgroup of $SU(2)$. This
additional invariance has to be fixed by means of a suitable gauge
condition on the diagonal component $A_{\mu }$, which will be chosen
to be of the Landau type, also adopted in lattice simulations
\cite{Amemiya:1998jz,Bornyakov:2003ee}, namely
\begin{equation}
\partial _{\mu }A_{\mu }=0\;.  \label{dgauge}
\end{equation}

The MAG is interesting from the lattice as well as continuum
viewpoint. It can be simulated numerically, as first discussed in
\cite{Kronfeld:1987ri,Kronfeld:1987vd}, while the MAG is also
renormalizable in the continuum.  Although strictly speaking, the
MAG as defined by (\ref{offgauge}) is not renormalizable. A slight
generalization has to be considered containing a gauge parameter
$\alpha$. Moreover, due to the nonlinearity of the gauge condition
(\ref{offgauge}), a quartic ghost interaction has to introduced in
the MAG action \cite{Min:1985bx}. The complete action turns out to
be given by
\begin{equation}
S=S_{\mathrm{YM}}+S_{\mathrm{MAG}}+S_{\mathrm{diag}}\;,
\label{smag}
\end{equation}
where $S_{\mathrm{MAG}},$\ $S_{\mathrm{diag}}$ are the gauge fixing
terms corresponding to the off-diagonal and diagonal sectors,
respectively, given by
\begin{widetext}
\begin{eqnarray}
S_{\mathrm{MAG}} &=&s\,\int d^{4}x\,\left( \overline{c}^{a}\left(
D_{\mu
}^{ab}A_{\mu }^{b}+\frac{\alpha }{2}b^{a}\right) -\frac{\alpha }{2}%
g\varepsilon \,^{ab}\overline{c}^{a}\overline{c}^{b}c\right) \;  \nonumber \\
&=&\int d^{4}x\left( b^{a}\left( D_{\mu }^{ab}A_{\mu }^{b}+\frac{\alpha }{2}%
b^{a}\right) +\overline{c}^{a}D_{\mu }^{ab}D_{\mu }^{bc}c^{c}+g\overline{c}%
^{a}\varepsilon ^{ab}\left( D_{\mu }^{bc}A_{\mu }^{c}\right) c
-\alpha g\varepsilon ^{ab}b^{a}\overline{c}^{b}c-g^{2}\varepsilon
^{ab}\varepsilon ^{cd}\overline{c}^{a}c^{d}A_{\mu }^{b}A_{\mu }^{c}-\frac{%
\alpha }{4}g^{2}\varepsilon ^{ab}\varepsilon ^{cd}\overline{c}^{a}\overline{c%
}^{b}c^{c}c^{d}\right) \;,  \label{smm}\nonumber\\
S_{\mathrm{diag}}&=&s\,\int d^{4}x\,\;\overline{c}\partial _{\mu
}A_{\mu }\;=\int d^{4}x\,\;\left( b\partial _{\mu }A_{\mu
}+\overline{c}\partial _{\mu }\left( \partial _{\mu }c+g\varepsilon
^{ab}A_{\mu }^{a}c^{b}\right) \right) \;.  \label{sgfd}
\end{eqnarray}
\end{widetext}
where $s$ is the nilpotent ($s^2=0$) BRST transformation, defined
through
\begin{eqnarray}
sA_{\mu }^{a} &=&-\left( D_{\mu }^{ab}c^{b}+g\varepsilon
^{\,ab}A_{\mu
}^{b}c\right) \;,  \nonumber \\
sA_{\mu } &=&-\left( \partial _{\mu }c+g\varepsilon ^{ab}A_{\mu
}^{a}c^{b}\right) \;,  \nonumber \\
sc^{a} &=&g\varepsilon \,^{ab}c^{b}c\;,\;\;\;\;sc =\frac{g}{2}\,\varepsilon \,^{ab}c^{a}c^{b},  \nonumber \\
s\overline{c}^{a} &=&b^{a}\;,\;\;\;\;s\overline{c} =b\;,  \nonumber \\
sb^{a} &=&0\;,\;\;\;\; sb =0\;,  \label{brs}\nonumber\\
sS&=&0\;.
\end{eqnarray}
$\left( \overline{c}^{a},c^{a}\right) $ and $\left( \overline{c}%
,c\right) $ are the off-diagonal and diagonal Faddeev-Popov ghosts,
while $\left( b^{a},b\right)$ denote the Lagrange multipliers. The
action (\ref{smag}) is renormalizable to all orders of perturbation
theory \cite{Fazio:2001rm,Dudal:2004rx}. Only at the end, when the
ultraviolet divergences are consistently treated and the theory is
renormalized, can one consider the limit $\alpha\rightarrow0$ which
is formally equivalent with the condition (\ref{offgauge}). We refer
to \cite{Dudal:2004rx,Fazio:2001rm} for details concerning the MAG
renormalization.

\subsection{Abelian dominance}
According to the concept of Abelian dominance, the low energy regime
of QCD can be expressed solely in terms of Abelian degrees of
freedom \cite{Ezawa:1982bf}. Lattice confirmations of the Abelian
dominance can be found in \cite{Suzuki:1989gp,Hioki:1991ai}. An
argument that can be interpreted in favour of it, is the fact that
the off-diagonal gluons would attain a large, dynamical mass. At
energies below the scale set by this mass, the off-diagonal gluons
should decouple, and in this way one should end up with an Abelian
theory at low energies.

\subsection{Why study propagators?}
The reader might question why one would study the propagators, which
are not even gauge invariant objects. Still, a certain number of
considerations can be outlined. To some extent, propagators are the
simplest Green's functions to be evaluated. Nowadays, it is possible
to obtain analytic estimates of the influence of nonperturbative
effects, like Gribov copies and condensates, on the infrared
behaviour of the propagators. As far as $4D$ Yang-Mills theory is
considered, this task appears to be very difficult for more
complicated Green's functions. In the last decade there has been an
intensive activity from the lattice community in the study of the
gluon and ghost propagators in a variety of gauges. We can thus
compare our theoretical predictions with the available lattice data.
So far, a certain number of gauges have been considered extensively
from theoretical as well as from the lattice point of view. This is
the case of the Landau, Coulomb and maximal Abelian gauge. In the
following, we shall focus on the MAG. We shall see that the
agreement between lattice results and theoretical investigations can
be considered satisfactory.

The study of the propagators might also provide a useful framework
to investigate the behaviour of the running coupling constant in the
infrared. This is the case, for example, of the Landau gauge, for
which the following relation holds
\begin{equation}
 Z_{g}=Z_{A}^{-1/2} Z_{c}^{-1}\;,
\end{equation}
to all orders of perturbation theory (see e.g.
\cite{Piguet:1995er}). If this relation is expected to be valid at
the nonperturbative level, the infrared behavior of the running
coupling constant $\alpha(p^2)$ could be investigated by looking at
the form factors of the gluon and ghost propagators. Such lattice
simulations and studies based on the Schwinger-Dyson equations have
provided evidence of the existence of an infrared fixed point in the
Landau gauge for a renormalization group invariant coupling constant
based on the gluon and ghost propagators (see e.g.
\cite{Lerche:2002ep}), i.e.
\begin{equation}
\alpha(0) \approx \frac{8.92}{N}\;,
\end{equation}
a result that has received some numerical confirmation too, see e.g.
\cite{Bloch:2003sk}.

Analogous relationships can be derived in the Coulomb gauge. A nice
updated work on the status of the running coupling constant in the
Coulomb gauge and its relationship with the Landau gauge can be
found in \cite{Fischer:2005qe}.

Interestingly, it turns out that also in the MAG, the
renormalization of the gauge coupling is related to that of the
fields, according to
\begin{equation}\label{drel}
Z_{g}=Z_{A_{\mathrm{diag}}}^{-1/2}\;,
\end{equation}
where $Z_{A_{\mathrm{diag}}}$ is the renormalization factor of the
diagonal component of the gauge field
\cite{Fazio:2001rm,Dudal:2004rx,Gracey:2005vu}. Analogously to the
Landau gauge, this relationship suggests that the infrared behavior
of the gauge coupling might be investigated by looking at the
diagonal gluon propagator.

\subsection{Lattice data in the MAG}
Lattice simulations of the MAG have given strong indications that
the off-diagonal gluons acquire a relatively large mass. Let us
mention here that the MAG condition (\ref{offgauge}) amounts to keep
as minimally as possible the off-diagonal components of the gauge
fields. As a consequence, one expects that the Abelian components
will play a predominant role. This is precisely the idea which
underlies the Abelian dominance hypothesis.

The first study of the gluon propagator on the lattice in the
maximal Abelian gauge was made in \cite{Amemiya:1998jz} in the case
of $SU(2)$. The gluon propagator was analyzed in coordinate space.
The off-diagonal component of the gluon propagator was found to be
short-ranged, exhibiting a Yukawa type behavior
\begin{equation}\label{yuk1}
G_{\mathrm{off}} \sim \frac{e^{-M_{\mathrm{off}}r}}{r^{3/2}}
\textrm{ with } M_{\mathrm{off}} \sim 1.2\mathrm{GeV}\;.
\end{equation}
The diagonal propagator was found to propagate over larger
distances.

More recently, a numerical investigation of the gluon propagator in
the maximal Abelian gauge has been worked out in
\cite{Bornyakov:2003ee}. Here, the gluon propagator was investigated
in momentum space. At low momenta, the diagonal component of the
gluon propagator has been found to be much larger than the
off-diagonal one. In particular, a Gribov like fit,
\begin{equation}
G_{\mathrm{diag}}(q)=\frac{q^{2}}{q^{4}+m_{dg}^{4}}\;, \label{gfit}
\end{equation}
turns out to be suitable for the diagonal component of the gluon
propagator. For off-diagonal gluons, a Yukawa type fit
\begin{equation}\label{gfit2}
G_{\mathrm{off}}(q)=\frac{1}{q^{2}+M_{\mathrm{off}}^{2}}\;,
\end{equation}
seems to work quite well. The mass parameter $M_{\mathrm{off}}$
appearing in the Yukawa fit is two times bigger that the
corresponding mass parameter $m_{\mathrm{diag}}$ of the Gribov fit,
namely
\begin{equation}
M_{\mathrm{off}}\approx 2m_{\mathrm{diag}}\;,
\end{equation}
where $M_{\mathrm{off}}\approx 1.2\mathrm{GeV}$, in agreement with
the result obtained in \cite{Amemiya:1998jz}. This implies that the
off-diagonal propagator is short-ranged as compared to the diagonal
one, a fact which can serve as an indication for a kind of Abelian
dominance.

We hope that it has become clear by now that the MAG is an important
gauge, and deserves our attention.

\section{Gribov copies in the MAG and restriction to the first Gribov horizon}
The Gribov ambiguity is due to the fact that a gauge fixing should
in principle select a \emph{single} representative of a gauge orbit
of any given gauge field configuration. Gribov has shown in the
seminal paper \cite{Gribov:1977wm} that, at least in the Landau and
Coulomb gauge, there exist gauge equivalent field configurations
obeying the Landau (or Coulomb) gauge. Gribov has worked out a
method to restrict more tightly the domain of integration in the
path integral for what considers the gauge fields.

Let us work out the condition for the existence of Gribov copies in
the MAG. In the case of small gauge transformations, this is easily
obtained by requiring that the gauge transformed fields
(\ref{gauge}), fulfill the same gauge conditions obeyed by $\left(
A_{\mu },A_{\mu }^{a}\right) $, i.e. (\ref{offgauge}) and
(\ref{dgauge}). Thus, to
the first order in the gauge parameters $\left( \omega ,\omega ^{a}\right) $%
, one gets
\begin{eqnarray}
-D_{\mu }^{ab}D_{\mu }^{bc}\omega ^{c}-g\varepsilon ^{bc}D_{\mu
}^{ab}\left( A_{\mu }^{c}\omega \right) &+&g\varepsilon ^{ab}A_{\mu
}^{b}\partial _{\mu }\omega\nonumber\\ +g^{2}\varepsilon
^{ab}\varepsilon ^{cd}A_{\mu }^{b}A_{\mu
}^{c}\omega ^{d} &=&0\;,  \label{offcopies0} \\
-\partial ^{2}\omega -g\varepsilon ^{ab}\partial _{\mu }\left(
A_{\mu }^{a}\omega ^{b}\right)  &=&0\;,  \label{diagcopies0}
\end{eqnarray}
which, due to (\ref{offgauge}) and (\ref{dgauge}) simplify to
\begin{eqnarray}
\mathcal{M}^{ab}\omega ^{b} &=&0\;,  \label{off1} \\
-\partial ^{2}\omega -g\varepsilon ^{ab}\partial _{\mu }\left(
A_{\mu }^{a}\omega ^{b}\right)  &=&0\;,  \label{de}
\end{eqnarray}
with
\begin{equation}
\mathcal{M}^{ab}=-D_{\mu }^{ac}D_{\mu }^{cb}-g^{2}\varepsilon
^{ac}\varepsilon ^{bd}A_{\mu }^{c}A_{\mu }^{d}\;.  \label{offop}
\end{equation}
This operator $\mathcal{M}^{ab}$ is recognized to be the
Faddeev-Popov operator for the off-diagonal ghost sector, see
\cite{Quandt:1997rg,Capri:2005tj}. It enjoys the property of being
Hermitian \cite{Bruckmann:2000xd}, thus having real eigenvalues.

One may thus expect the appearance of Gribov copies in the MAG. And
indeed, a normalizable zero mode of the Faddeev-Popov operator
(\ref{offop}) was constructed in \cite{Bruckmann:2000xd}.

It was shown in \cite{Capri:2005tj} that the partition function for
the MAG, described by the action (\ref{smag}) can be recast into the
form
\begin{equation}
\mathcal{Z}=\int DA_{\mu }DA_{\mu }^{a}\;\delta \left( D_{\mu
}^{ab}A_{\mu }^{b}\right) \delta \left( \partial _{\mu }A_{\mu
}\right) \det \mathcal{M}^{ab} e^{-S_{\mathrm{YM}}\;}\;,
\label{pftf}
\end{equation}
in the limit $\alpha\rightarrow 0$ and after integration over the
Lagrange multipliers as well as over the off-diagonal and diagonal
ghost fields, in the latter case a nontrivial field transformation
was used.

Let us now briefly explain the idea of Gribov to restrict the domain
of integration further \cite{Gribov:1977wm}, applied to the MAG. The
interested reader can find the details for the MAG in
\cite{Capri:2005tj}. We define
the Gribov region $\mathcal{C}%
_{0}$ as the set of fields fulfilling the gauge conditions (\ref
{offgauge}), (\ref{dgauge}) and for which the Faddeev-Popov operator $%
\mathcal{M}^{ab}$ is positive definite, namely
\begin{equation}
\mathcal{C}_{0}=\left\{ A_{\mu },\;A_{\mu }^{a},\;\partial _{\mu
}A_{\mu }=0,\;D_{\mu }^{ab}A_{\mu
}^{b}=0,\;\mathcal{M}^{ab}>0\right\} \;.  \label{gr}
\end{equation}
The boundary, $l_{1}$, of the region $\mathcal{C}_{0}$, where the
first vanishing eigenvalue of $\mathcal{M}^{ab}$ appears, is called
the first Gribov horizon. The restriction of the domain of
integration to this region is supported by the possibility of
generalizing to the maximal Abelian gauge Gribov's original result
\cite{Gribov:1977wm} stating that for any field located near a
horizon there is a gauge copy, close to the same horizon, located on
the other side of the horizon. This statement for the MAG was
explicitly proven in \cite{Capri:2005tj}.

The idea of Gribov was now to restrict the domain of integration to
the Gribov region. Therefore, we modify the MAG partition function
(\ref{pftf}) to
\begin{equation}
\mathcal{Z}=\int DA_{\mu }^{a}DA_{\mu }\;\det \mathcal{M}%
^{ab}(A)\;\delta \left( D_{\mu }^{ab}A_{\mu }^{b}\right) \delta
\left( \partial _{\mu }A_{\mu }\right) e^{-S_{YM}}\mathcal{V}(\mathcal{C}%
_{0})\;,  \label{pf}
\end{equation}
where the factor $\mathcal{V}(\mathcal{C}_{0})$ implements the
restriction to the region $\mathcal{C}_{0}$. Following
\cite{Gribov:1977wm}, the factor $ \mathcal{V}(\mathcal{C}_{0})$ can
be accommodated for by means of a so-called ``no pole condition'' on
the off-diagonal ghost two-point function, which is nothing else
than the inverse of the Faddeev-Popov operator $\mathcal{M}^{ab}$.
More precisely, denoting
by $\mathcal{G}(k,A)$ the Fourier transform of $\left( \mathcal{M}%
^{ab}\right) ^{-1}$, we shall require that $\mathcal{G}\left(
k,A\right) $ has no poles for a given nonvanishing value of the
momentum $k$, except for a singularity at $k=0$, corresponding to
the boundary of $\mathcal{C}_{0}$, i.e. to the first Gribov horizon
$l_{1}$ \cite{Gribov:1977wm}. This no pole
condition can be easily understood by observing that, within the region $%
\mathcal{C}_{0}$, the Faddeev-Popov operator $\mathcal{M}^{ab}$ is
positive definite. This implies that its inverse, $\left(
\mathcal{M}^{ab}\right) ^{-1}$, and thus the Green function
$\mathcal{G}(k,A)$, can become large only when approaching the
horizon $l_{1}$, where the operator $\mathcal{M}^{ab}$ has a zero
mode.

\subsection{Effects on the propagators}
The explicit implementation of the factor
$\mathcal{V}(\mathcal{C}_0)$ can be found in \cite{Capri:2005tj}.
Currently, we are more interested in the eventual implications on
the propagators of the $SU(2)$ MAG theory.

For off-diagonal gluon propagator, it was found that
\begin{equation}
\left\langle A_{\mu }^{a}A_{\nu }^{b}\right\rangle_q =\delta ^{ab}\frac{%
1}{q^{2}}\left( \delta _{\mu \nu }-\frac{q_{\mu }q_{\nu
}}{q^{2}}\right) \;, \label{offdp}
\end{equation}
while for the diagonal gluon propagator, it holds that
\begin{equation}
\left\langle A_{\mu }A_{\nu }\right\rangle_q
=\frac{q^{2}}{q^{4}+\gamma ^{4}}\left( \delta _{\mu \nu
}-\frac{q_{\mu }q_{\nu }}{q^{2}}\right) \;. \label{dprop}
\end{equation}
We notice the appearance of the so-called Gribov mass parameter,
which is the solution of the gap equation
\begin{equation}
\frac{3}{4}g^{2}\int \frac{d^{4}p}{\left( 2\pi \right) ^{4}}\frac{1}{%
p^{4}+\gamma ^{4}}=1\;.  \label{gapmag}
\end{equation}
One sees that the diagonal component, (\ref{dprop}), is suppressed
in the infrared, exhibiting the characteristic Gribov type behavior.
The off-diagonal components, (\ref {offdp}), remains unchanged.
However, as we shall soon see, its infrared behavior turns out to be
modified once the gluon condensate $\left\langle A_{\mu }^{a}A_{\mu
}^{a}\right\rangle $ is taken into account.

Concerning the ghost propagators, it can be shown that the diagonal
ghost propagator is left unaffected by the restriction to the first
Gribov region (see \cite{Capri:2005tj}). For the (trace of the)
off-diagonal ghost propagator, one finds
\begin{equation}
\mathcal{G}(q)_{q\approx 0}\approx \frac{\gamma ^{2}}{q^{4}}\;,
\label{offgh}
\end{equation}
exhibiting the typical infrared enhancement due to the Gribov
restriction.

\section{Dynamical off-diagonal gluon mass}
In \cite{Dudal:2004rx}, the generation of the dimension two gluon
condensate $\left\langle A_\mu^a A_\mu^a\right\rangle$ was
discussed. A renormalizable effective potential for $\left\langle
A_{\mu }^{a}A_{\mu }^{a}\right\rangle $ in the MAG has been
constructed and evaluated in analytic form in \cite{Dudal:2004rx}. A
nonvanishing condensate $\left\langle A_{\mu }^{a}A_{\mu
}^{a}\right\rangle $ is favoured since it lowers the vacuum energy.
As a consequence, a dynamical tree level mass for
off-diagonal gluons is generated. The combination of the condensate $%
\left\langle A_{\mu }^{a}A_{\mu }^{a}\right\rangle $ within the
Gribov approximation can be performed along the lines outlined in
\cite{Sobreiro:2004us}, where the effects of the Gribov copies on
the gluon and ghost propagators in the presence of the dimension two
gluon condensate have been worked out in the Landau gauge. Following
\cite{Dudal:2004rx}, the dynamical mass generation is accounted for
by adding to the gauge fixed Yang-Mills action the following term
\footnote{We recall here that in principle, one has to employ the
renormalizable action (\ref{smag}). As a consequence, the slightly
more general operator $\frac{1}{2}A_\mu^a
A_\mu^a+\alpha\overline{c}^a c^a$ has to used. At the end, one can
consider the limit $\alpha\rightarrow0$. We refer to
\cite{Dudal:2004rx} for more details.}
\begin{equation}
S_{\sigma }=\int d^{4}x\left( \frac{\sigma ^{2}}{2g^{2}\zeta }+\frac{1}{2}%
\frac{\sigma }{g\zeta }A_{\mu }^{a}A_{\mu }^{a}+\frac{1}{8\zeta
}\left( A_{\mu }^{a}A_{\mu }^{a}\right) ^{2}\right) \;.  \label{m2}
\end{equation}
The field $\sigma $ is an auxiliary field which allows one to study
the condensation of the local operator $A_{\mu }^{a}A_{\mu }^{a}$,
since \cite{Dudal:2004rx}
\begin{equation} \left\langle \sigma
\right\rangle =-\frac{g}{2}\left\langle A_{\mu }^{a}A_{\mu
}^{a}\right\rangle \;.  \label{m3}
\end{equation}
The dimensionless parameter $\zeta $ in expression $\left(
\ref{m2}\right) $ is needed to account for the ultraviolet
divergences present in the vacuum correlation function $\left\langle
A^{2}(x)A^{2}(y)\right\rangle $. For the details of the
renormalizability properties of the local operator $A_{\mu
}^{a}A_{\mu }^{a}$ in the maximal Abelian gauge we refer to
\cite{Dudal:2004rx} and references therein. The inclusion of the
term $S_{\sigma }$ is the starting point for evaluating the
renormalizable effective potential $V(\sigma )$ for the auxiliary
field $\sigma $, which is moreover consistent with the
renormalization group equations. The minimum of $V(\sigma )$ occurs
for a nonvanishing vacuum expectation value $\sigma$, $\left\langle
\sigma \right\rangle \neq 0$. In particular, the first order
off-diagonal dynamical gluon mass turns out to be
\cite{Dudal:2004rx}
\begin{equation}
m^{2}\equiv\frac{\left\langle \sigma \right\rangle }{g\zeta }\approx
(2.25\lms)^2\;. \label{msig}
\end{equation}
The inclusion of the action $S_{\sigma }$ leads to a partition
function which is still plagued by the Gribov copies. It might be
useful to note in fact that $S_{\sigma }$ is left invariant by the
local gauge transformations (\ref{gauge}), supplemented with
\begin{eqnarray}
\delta \sigma &=&gA_{\mu }^{a}D_{\mu }^{ab}{\omega }^{b}\;.
\label{str}
\end{eqnarray}
The same procedure as before in the absence of $\left\langle A_\mu^a
A_\mu^a\right\rangle$ can be repeated, and the outcome is that
(\ref{dprop}) and (\ref{offgh}) are still valid, but the
off-diagonal gluon propagator (\ref{offdp}) gets modified to
\begin{equation}
\left\langle A_{\mu }^{a}A_{\nu }^{b}\right\rangle_q =\delta ^{ab}\frac{%
1}{q^{2}+m^2}\left( \delta _{\mu \nu }-\frac{q_{\mu }q_{\nu
}}{q^{2}}\right) \;, \label{offdpbis}
\end{equation}
at lowest order.

\section{Comparison with MAG lattice data}
Although the extrapolation of the lattice data in the region
$q\approx 0$ is a difficult task, requiring large lattice volumes,
the results (\ref{dprop}) and (\ref{offdpbis}) on the (transverse)
diagonal and off-diagonal components of the gluon propagator can be
considered to be in qualitative agreement with the lattice results,
especially with the two parameter fits (\ref{gfit}) and
(\ref{gfit2}). Concerning now the ghost propagator, to our
knowledge, no lattice data is available so far.
\section{Conclusion}
We have reviewed some analytically obtained results on the infrared
behavior of the gluon and ghost propagators in the MAG. The diagonal
gluon propagator displays a Gribov type behavior (\ref{dprop}),
while the off-diagonal one has a Yukawa type behavior
(\ref{offdpbis}), with an off-diagonal mass originating from the
dimension two condensate $\langle A_{\mu}^{a}A_{\mu}^{a} \rangle$.
These results are in satisfactory agreement with the available
lattice data from \cite{Amemiya:1998jz,Bornyakov:2003ee}, summarized
in (\ref{yuk1}), (\ref{gfit}) and (\ref{gfit2}).

We hope these results might stimulate further investigation of
Yang-Mills theories in the MAG. A first point of interest is the
behaviour of the ghost propagators in the infrared, on which there
is no lattice data available yet. A look at the off-diagonal ghost
propagator from lattice simulations would be of a certain interest
in order to improve our understanding of the influence of the Gribov
copies in the MAG. A second point worth to be investigated is the
all order relation (\ref{drel}), which could allow one to get
information on the behavior of the gauge coupling constant in the
infrared from the study of the diagonal component of the gluon
propagator, allowing thus for a comparison with similar results
obtained in Landau and Coulomb gauges.

\section*{Acknowledgments}
\noindent S.P.~Sorella would like to thank the ``I
LAWHEP''-organizers for this interesting conference. The Conselho
Nacional de Desenvolvimento Cient\'{i}fico e Tecnol\'{o}gico
(CNPq-Brazil), the Faperj, Funda{\c{c}}{\~{a}}o de Amparo {\`{a}}
Pesquisa do Estado do Rio de Janeiro, the SR2-UERJ and the
Coordena{\c{c}}{\~{a}}o de Aperfei{\c{c}}oamento de Pessoal de
N{\'\i}vel Superior (CAPES) are gratefully acknowledged for
financial support. D.~Dudal is a postdoctoral fellow of the
\emph{Special Research Fund} of Ghent University.

\end{document}